\documentclass[prd,twocolumn,showpacs,amsmath,amssymb]{revtex4}
\usepackage{graphicx}
\usepackage{dcolumn}
\usepackage{bm}

\newcommand{\EE}{e^+e^-}
\newcommand{\psp}{\psi^{\prime}}

\newcommand{\jpsi}{J/\psi}
\newcommand{\pp}{\pi^+\pi^-}
\newcommand{\kk}{K^+K^-}
\newcommand{\kskl}{K^0_S K^0_L}
\newcommand{\Ctt}{C_{\theta}}
\newcommand{\Stt}{S_{\theta}}
\newcommand{\beq}{\begin{equation}}
\newcommand{\eeq}{\end{equation}}
\def\eref#1{(\ref{#1})}
\def\Journal#1#2#3#4{{#1} {\bf #2}, #3 (#4)}

\def\PRD{Phys. Rev. D}

\begin{document}

\title{Electromagnetic cross section of charmonium decay}

\author{X.H.Mo$^{1,2}$
\\  \vspace{0.2cm} {\it
$^{1}$ Institute of High Energy Physics, CAS, Beijing 100049, China\\
$^{2}$ University of Chinese Academy of Sciences, Beijing 100049, China\\
}
}
\email{moxh@ihep.ac.cn}
\date{\today}
	
\begin{abstract}
As an important measurable quantity, the electromagnetic cross sections of charmonium decay in an $\EE$ collider are valuated for fifteen final states. A general scheme of $lS$-coupling analysis is adopted for constructing covariant helicity-coupling amplitudes involving two-body decays with arbitrary integer spins. Based on the decay rate, the collision cross section is obtained finally. A number of illustrative examples is worked out to expound and incarnate the general formulas.
\end{abstract}
\maketitle

\section{Introduction}
Although the Standard Model (SM) has been accepted as a universally appreciated theory basis in the high energy community, it is still hard to calculate the wanted experimental observable from the first principle of the SM for a great many of processes, especially when the strong interaction is involved. Quantum chromodynamics (QCD) as a widely appreciated theory of strong interaction, has been proved to be very successful at high energy when the calculation can be executed perturbatively. However, its validity at the nonperturbative regime, such as $\jpsi$ and $\psp$ resonance regions, needs more experimental guidance.

At present, the phenomenological analysis plays an important role for deepening our understanding of the general principle. As a matter of fact, many phenomenological models are constructed for charmonium decay~\cite{Kowalski:1976mc}-\cite{moxh2025},
the parametrization of various decay modes are obtained, such as the pseudoscalar and pseudoscalar mesons (PP), vector and pseudoscalar mesons (VP), octet baryon-pair, and so on.

Recently, a systematical parametrization scheme has been proposed~\cite{moxh2023,moxh2024,moxh2025}, by virtue of $SU(3)$ flavor symmetry, the effective interaction Hamiltonian in tensor form is obtained according to group representation theory. In the light of flavor singlet principle, a systematical parametrization is realized for all charmonium decays, involving binary decays, ternary decays and radiative decays. The parameters appeared in the parametrization scheme, which can be figured out by virtue of data analysis, will shed the light on charmonium decay dynamics. Nevertheless, for data analysis the collision cross section is the crucial bridge between the phenomenological model and the experimental information. Moreover, the differential cross section that contains the angular distribution of particles furnishes a pragmatical tool for analyzing the data systematically and comprehensively.

This paper is devoted to the calculation of the electromagnetic cross sections of charmonium decay. In the next section, the decay rate is expressed by
covariant helicity-coupling amplitudes involving two-body decays with arbitrary integer spins. A general scheme of $lS$-coupling analysis is developed to construct the Lorentz invariant amplitude and lots of illustrative examples are worked out to explained the general formulas. The section that follows focuses on the calculation of the electromagnetic cross sections of charmonium decay. The relation between the decay rate and the collision cross section is derived in this section. After that is a summary section. Relegated to the appendix are materials on various calculation details.

\section{Decay rates}\label{xct_dkrate}
The helicity formalism is more general since it is valid for any mass and spin of particles involved in the decay. Such a formalism has been established by Suh-Urk Chung in a series of papers~\cite{chungsu2008,chungsu1998,chungsu1993,chungsu1971}. Here we only recapitulate the main ideas and collect the formulas needed for the following calculations.

Consider a two-body decay $V \to X+Y$, where parent particle with spin $J$ is rest and daughter particles with spins $s$ and $\sigma$ respectively fly out back-to-back. See Table~\ref{talforvxy} for all other relevant quantities.

\begin{table}[hbt]
\caption{\label{talforvxy} Two-body decay: $V (J) \to X (s)+ Y (\sigma)$.} \center
\begin{tabular}{lccc}\hline \hline
Quantity      & Parent         &    Daughter 1  & Daughter 2  \\ \hline
Spin          & $J$            & $s$            & $\sigma$ \\
Parity        & $\eta_J$       & $\eta_s$       & $\eta_\sigma$  \\
Helicity      &                & $\lambda$      & $\nu $ \\
Momentum      & ${\bf p}$      & ${\bf q}$      & ${\bf k}$ \\
Energy        & $ p_0$         & $ q_0$         & $ k_0$  \\
Mass          & $w$            & $w_s$          & $w_\sigma$  \\
Energy/mass   &                & $\gamma_s$     & $\gamma_\sigma$  \\
Velocity      &                & $\beta_s$      & $\beta_\sigma$  \\
Wave function & $\phi^{*} (\lambda-\nu)$  &  $\omega (\lambda)$ & $\epsilon (\nu)$    \\
 \hline \hline
\end{tabular}
\end{table}

The key idea is as follows: there exist two ket states $|S m_S \rangle$ and $|l m_l \rangle$ in the decay process $V (J) \to X (s)+ Y (\sigma)$. They act like the states at rest in the rest frame of parent particle (PPRF), so there exist three at-rest states in the problem; $|J M \rangle$ in the initial system and  $|l m_l \rangle$ in the final system, at the rest of PPRF. The third state $|S m_S \rangle$ acts as if it is again at the rest of PPRF. It can proved~\cite{chungsu2008} that $J$, $S$, and $l$ are rotational invariants, moreover, $\hat{J}=\hat{S}+\hat{l}$, then the operator $\hat{l}$ and $\hat{S}$ provide a fully prescription for the conventional $lS$ coupling scheme for two-particle final states.

The polarization four-vectors or wave functions appropriate for the particles ``$J=1$'',  ``$s=1$'', and ``$\sigma=1$'' are well known. In the particle rest frame (PRF),

\beq
\left.\begin{array}{rl}
p^\alpha = & (w; 0,0,0)~,    \\
q^\alpha = & (q_0; 0,0,q) = (\gamma_s w_s; 0,0,\gamma_s \beta_s w_s)~, \\
k^\alpha = & (k_0; 0,0,-q) = (\gamma_\sigma w_\sigma; 0,0,-\gamma_\sigma \beta_\sigma w_\sigma)~, \\
r^\alpha = & (q_0-k_0; 0,0,2q)~,
\end{array}\right.~~
\label{prfvriables}
\eeq
where $w=q_0+k_0$, $q_0=\sqrt{w_s^2+q^2}$,  $k_0=\sqrt{w_\sigma^2+q^2}$, and $r=q-k$, the wave functions are given by

\beq
\left.\begin{array}{rl}
\phi^\alpha(\pm) = & {\displaystyle \mp \frac{1}{\sqrt{2}} (0; 1,\pm i,0) }~,    \\
\phi^\alpha(0) = & {\displaystyle (0; 0, 0, 1) }~,    \\
\omega^\alpha(\pm) = & {\displaystyle \mp \frac{1}{\sqrt{2}} (0; 1,\pm i,0) }~,    \\
\omega^\alpha(0) = & {\displaystyle (\gamma_s \beta_s; 0, 0, \gamma_s ) }~,    \\
\epsilon^\alpha(\pm) = & {\displaystyle \mp \frac{1}{\sqrt{2}} (0; 1,\pm i,0) }~,    \\
\epsilon^\alpha(0) = & {\displaystyle (-\gamma_\sigma \beta_\sigma; 0, 0, \gamma_\sigma ) }~.
\end{array}\right.~~
\label{prfwvfns}
\eeq
Note that
$$ p_\alpha \phi^\alpha(\lambda) = q_\alpha \omega^\alpha(\lambda) = k_\alpha \epsilon^\alpha(\lambda) =0$$
for any $\lambda$.

These polarization four-vectors satisfy
\beq
\left.\begin{array}{c}
 p_\alpha \phi^\alpha(m) = 0, ~~~ \phi^{*}_\alpha(m)  \phi^\alpha(m^{\prime}) = -\delta_{m m^{\prime}}~,    \\
\sum \phi_\alpha(m) \phi^{*}_\beta (m)= \tilde{g}_{\alpha \beta} (w) ~,
\end{array}\right.~~
\label{rlnofwvfns}
\eeq
where
\beq
\tilde{g}_{\alpha \beta} (w)=- g_{\alpha \beta} +\frac{p_\alpha p_\beta}{w^2}~.
\label{defofgab}
\eeq
The last equality of Eq.~\eref{rlnofwvfns} is the usual projection operator for spin-1 states~\cite{chungsu1998}.
$\omega$ and $\epsilon$ satisfy similar conditions with their own $\tilde{g}$'s.

The high spin state can be constructed from the low spin state step by step, and the general spin-$j$ wave function reads~\cite{chungsu1998}
\beq
\left.\begin{array}{rl}
\phi^{j}_{\delta_1 \cdots \delta_j}(m) &= {\displaystyle [a^{j} (m)]^{1/2} \sum_{m_0} 2^{m_0/2}}    \\
    & {\displaystyle \times \sum_{P} \underbrace{\phi_{\alpha_1} (+) \cdots}_{m_{+}}
    \underbrace{\phi_{\beta_1} (0) \cdots}_{m_{0}} \underbrace{\phi_{\gamma_1} (-) \cdots}_{m_{-}} }~,
\end{array}\right.~~
\label{spinjwvfn}
\eeq
where
\beq
a^{j} (m)= \frac{(j+m)! (j+m)!}{(2j)!}
\label{defofajm}
\eeq
and the indices $\{ \delta_1 \cdots \delta_j \}$ have been split out into three distinct sets in the second summation, i.e., $\{\alpha_i \}$ with ($i=1, \cdots, m_+$), $\{\beta_i \}$ with ($i=1, \cdots, m_0$), and $\{ \gamma_i \}$ with ($i=1, \cdots, m_-$), where $m_{\pm}$ stands for the number of $\phi(\pm)$'s and $m_{0}$ for $\phi(0)$'s. Note that
\beq
j = m_+ +m_0 +m_-
\label{defofj}
\eeq
and
\beq
m = m_+ - m_- ~,
\label{defofm}
\eeq
in addition,
\beq
2 m_{\pm} = j \pm m - m_0 ~,
\label{defofmpm}
\eeq
which indicates that the right-hand side must be always even. The first sum in Eq.~\eref{spinjwvfn} goes over the allowed values of $m_0$ given $j$ and $m$. It is clear that the maximum is given by $j-m$, so that $m_0$ ranges from $0(1),~2(3),~\cdots$ to $j-m=$even(odd). The second sum in Eq.~\eref{spinjwvfn} represents a summation on the permutations
$$ \{~ (+)(+) \cdots~(0)(0) \cdots~(-)(-) \cdots ~\}~.$$
It is seen readily that the number of terms in the summation is
\beq
b^{j} (m,m_0)= \frac{j!}{ m_+! m_0! m_-!}~.
\label{defofbjm}
\eeq

The wave function for a two-body decay with the synthetic intrinsic spin $S$ is formed out of the $s$ and $\sigma$ polarization four-vectors
\beq
\left.\begin{array}{rl}
\psi(s,\sigma; S m_S) &= {\displaystyle \sum_{m_s m_{\sigma}} \langle s m_s \sigma m_{\sigma} | S m_S \rangle } \\
  & \times \phi^{s}_{\alpha_1 \cdots \alpha_s} (m_s) \phi^{\sigma}_{\beta_1 \cdots \beta_\sigma} (m_{\sigma})~,
\end{array}\right.~~
\label{totspinwvfn}
\eeq
where $\langle s m_s \sigma m_{\sigma} | S m_S \rangle$ is the Clebsch-Gordan coefficient; rank-$s$ tensor $\phi^{s}(m_s)$ and rank-$\sigma$ tensor $\phi^{\sigma}(m_\sigma)$ are the spin $j=s,\sigma$ wave functions constructed in Eq.~\eref{spinjwvfn}. Consequently, $\psi(s,\sigma; S m_S)$ is a tensor of rank-($s+\sigma$).

The wave function $\chi(l m_l)$ for the orbital-angular momentum $l$ is the rank-$l$ tensor, just like $\phi^{j}_{\delta_1 \cdots \delta_j}(m)$ in Eq.~\eref{spinjwvfn}. As a matter of fact, one can simplify the treatment of orbital-angular momentum by defining $\chi(l m_l)$ along the helicity axis (that is $z$ axis), then for two-body decay, merely the collinear vectors along the $z$ axis need to be dealt with exclusively and the wave function $\chi(l m_l)$ comes with a zero $z$ component. This means that in the PRF the orbital-angular momentum does not have time component, therefore $m_l=0$,
\beq
\chi(l 0)= c_l \phi^{l}_{\delta_1 \cdots \delta_l}(0)~,~~~ c_l=\left[ \frac{2^l (l!)^2}{(2l)!} \right]^{1/2}~.
\label{defofchil0}
\eeq

Now the Lorentz invariant amplitude can be constructed as follows
\beq
A^J_{lS}(s,\sigma;\delta)= \left[ p^n, \psi(s,\sigma; S m_S), \chi(l 0), \phi^{*} (J \delta) \right]~.
\label{defofampa}
\eeq
This amplitude corresponds to the transition $| J \delta \rangle \to | S \delta \rangle + |l 0 \rangle$ in which the decay products are aligned along $z$ axis. Here $\delta = \lambda - \nu $, the wave function $\phi^{*} (J \delta)$ refers to the initial spin $J$ state or the parent particle. The value of $l$ will be determined by the $lS$-coupling recipe. First, the $S$ and $l$ are coupled according to the usual angular momentum coupling rule; second, only allowed is the $l$ when the coupled values of $S$ and $l$ contain $J$.

The notation $[\cdots]$ indicates Lorentz contraction of all of the relevant Lorentz indices in the PRF. The following rules are adopted for the contraction~:
\begin{enumerate}
\item with the modified metric $\tilde{g} (w)$ [see Eq.~\eref{defofgab} ] and only with this metric, if $s+\sigma + l -J =$even. Here $p^{\mu}$ never appears in the amplitudes, i.e., $n=0$.
\item with the totally antisymmetric rank-4 tensor ($\epsilon^{\alpha \beta \mu \nu}$) together with the four-momentum $p^{\mu}$, and with $\tilde{g} (w)$ whenever necessary, if $s+\sigma + l -J =$odd. The the usage of $\epsilon^{\alpha \beta \mu \nu}$ and $p^{\mu}$ is restricted to be once and only once, so that $n=1$.
\item for angular momentum $l$, only amplitudes proportional to $r^l$ are considered ($r^l$ rule).
\end{enumerate}
By virtue of these rules, the $A^J_{lS}(s,\sigma;\delta)$ can be figured out to be
\beq
\left.\begin{array}{rl}
A^J_{lS}(s,\sigma;\delta) &= {\displaystyle \left( \frac{2l+1}{2J+1} \right)^{1/2} \langle l 0 S \delta | J \delta\rangle \langle s \lambda \sigma -\nu |S \delta \rangle } \\
  & \times w^n r^l f^s_{\lambda} (\gamma_s) f^{\sigma}_{\nu} (\gamma_\sigma)~.
\end{array}\right.~~
\label{xpnofampa}
\eeq
The general formula for the $f$ function reads
\beq
f^j_{m} (\gamma) = a^{j} (m) \sum_{m_0} b^{j} (m,m_0) (2 \gamma)^{m_0}~,
\label{fmloffnf}
\eeq
where $j$, $m$, and $\gamma$ can stand for $s$, $\lambda$, and $\gamma_s$ or $\sigma$, $\nu$, and $\gamma_\sigma$, respectively.

The summation of all possible $l$ and $S$ yields the invariant helicity-amplitude
\beq
F^J_{\lambda \nu} = \sum_{l S} g_{l S} \cdot A^J_{lS}(s,\sigma;\delta)~,
\label{defofampf}
\eeq
where the complex parameters $g_{l S}$ are unknown, to be determined from experiments.

To pin down $F^J_{\lambda \nu}$, the following rules are to be satisfied~:
\begin{enumerate}
\item Sine $\delta=\lambda - \nu$ is the projection of $J$ along $z$ axis, so $|\lambda - \nu| \leqslant J$.
\item The product of intrinsic parities is linked to the allowed orbital-angular momentum via the relation $\eta_J=\eta_s \eta_\sigma (-)^l$.
\item Parity conservation in the decay leads to the relationship
\beq
F^J_{\lambda \nu} = \eta_J \eta_s \eta_\sigma (-)^{J-s-\sigma} F^J_{-\lambda -\nu} =(-)^{s+\sigma+l-J} F^J_{-\lambda -\nu}~.
\label{rlnforffn}
\eeq
\end{enumerate}

The effective use of these rules serves to reduce the number of independent amplitudes to roughly one-half of the nominal number. Precisely speaking, the number of independent $F^J_{\lambda \nu}$'s can be evaluated by the formula
\beq
N_J = (a+b+1)(a-s+\sigma +1)+(s+\sigma-a)(2J+1)~,
\label{nmbofampf}
\eeq
where
\beq
\left.\begin{array}{rl}
a &= \min \{ J, s+\sigma \} ~,\\
b &= \min \{ J, s-\sigma \} ~,
\end{array}\right.~~
\label{defofaandb}
\eeq
and one has assumed here that $s \geqslant \sigma $. Some numbers of $N_J$ for various final states are evaluated and summarized in Table~\ref{tabnmboff}.

\begin{table}[hbt]
\caption{\label{tabnmboff} The number of independent $F^J_{\lambda \nu}$'s. The symbols  $S, P, V, A$, and $T$ denote scalar, pseudoscalar, vector, axial vector, and tensor, respectively.} \center
\begin{tabular}{c|ccccc}\hline \hline
$N_J$  & $S$   & $P$   & $V$   & $A$   & $T$    \\ \hline
$M^{P}$& $0^+$ & $0^-$ & $1^-$ & $1^+$ & $2^+$  \\ \hline
$S$    & 1     & 0     & 2     & 1     & 2  \\
$P$    &       & 1     & 1     & 2     & 1  \\
$V$    &       &       & 4     & 3     & 5  \\
$A$    &       &       &       & 4     & 4  \\
$T$    &       &       &       &       & 7  \\  \hline \hline
\end{tabular}
\end{table}

Finally, we access to the decay rate~\cite{quangpham} for the process $V \to X+Y$
\beq
\frac{d\Gamma (V \to X+Y)}{d\Omega} = \frac{|{\cal M}^{\Gamma}|^2}{32 \pi^2} \cdot \frac{|q_f|}{m_V^2}~,
\label{decayrate}
\eeq
where $m_V$ is the mass of particle $V$ and $q_f$ is the momentum of either the $X$ or the $Y$. The decay amplitude reads
\beq
{\cal M}^{\Gamma}=\sum_{\lambda \nu} D^{J*}_{M \delta} (\varphi, \theta, 0) \cdot F^J_{\lambda \nu} ~,
\label{mtxmgamma}
\eeq
whence
\beq
|{\cal M}^{\Gamma}|^2=\sum_{\lambda \nu} |d^{J}_{M \delta}(\theta)|^2 \cdot |F^J_{\lambda \nu}|^2 ~,
\label{mtxsqmgamma}
\eeq
where $M$ is the spin projection of $J$, and
\beq
D^{j}_{m^{\prime} m} (\alpha, \beta, \gamma)= e^{-i m^{\prime} \alpha} d^{j}_{m^{\prime} m} e^{-i m \gamma} ~,
\label{rlnofdfn}
\eeq
is the usual $D$-function ($d$-function) in angular momentum theory~\cite{ross1957}.

A remark is in order here. In the light of Eq.~\eref{decayrate}, the $|{\cal M}^{\Gamma}|^2/M^2_V$ is a unitless quantity, this form is usually adopted for the evaluation of the decay amplitude. Such a choice leads to the requirement that $A^J_{lS}(s,\sigma;\delta)$ should be unitless as well. To this end, one choice is to normalize the momentum $p$ in Eq.~\eref{defofampa} by $w$, then the $w^n$ is to be disappeared from Eq.~\eref{xpnofampa}. As far as the barrier factor $r^l$ is concerned, the variable $r$ is measured in units of the scale factor $r_0$ (e.g. 1 fm), so that $r^l$ should be understood as (unit-free) quantity  $(r/r_0)^l$. Nevertheless, except for the evaluation of the decay amplitdue, the original symbols are kept without change.

In the following subsections, a few of examples are spell out to exhibit the application of the aforementioned rules. Generally speaking, in the rest frame of a massive parent particle, the angular distribution of the two-body decay is a trivial event distribution. However, the charmonium that we confront is produced from an $\EE$ collider, there are only two helicity projections, which leads to abundant situations. Since such processes are really concered for our study, the polarization due to $\EE$ production is left implicit henceforth.

\subsection{$1^{-}\rightarrow 0^{+} +  0^{-}$}
For this mode, $s,\sigma=0$ so that $\lambda,\nu=0$; in addition $J=1$, together with $\eta_J=-1, \eta_s=+1, \eta_\sigma=-1$, according to  Eq.~\eref{rlnforffn}, it is obtained immediately
$$ F^1_{00}=-F^1_{00}~,$$
that is $ F^1_{00}=0$. Therefore, this decay mode is forbidden by parity conservation. For an example, the decay like $\psi \to a_0(1450)\pi^0$ is forbidden.

\subsection{$1^{-}\rightarrow 0^{+} +  1^{-}$}
First, the restriction due to the orbital-angular momentum, $\eta_J=\eta_s \eta_\sigma (-)^l$, leads to $l$=even. Second, $S=s+\sigma=1$, then from $1 \oplus 0=1$; $1 \oplus 2=1,2,3$; and $1 \oplus 4=3,4,5$, so that $l$ can only be 0 or 2, that is only $S$-wave and $D$-wave are allowed. Third, Eq.~\eref{rlnforffn} yields
$$ F^1_{0,1}=F^1_{0,-1}~,$$
so that there are merely two independent invariant helicity-amplitudes, $F^1_{0,1}$ and $F^1_{0,0}$. Fourth, since $s+\sigma + l -J =$even, then $n=0$. By virtue of Eq.~\eref{xpnofampa},
\beq
\left.\begin{array}{rl}
A^1_{01}(\lambda) &= {\displaystyle \sqrt{\frac{1}{3}} r^0 f^{1}_{\lambda} (\gamma_s) }~,  \\
A^1_{21}(\lambda) &= {\displaystyle \sqrt{\frac{5}{3}} \langle 2 0 1 \lambda |1 \lambda \rangle  r^2 f^{1}_{\lambda} (\gamma_s) }~.
\end{array}\right.~~
\label{psisvampa}
\eeq
By virtue of Eq.~\eref{defofampf},
\beq
\left.\begin{array}{rcl}
F^{1}_{0,1}&=& {\displaystyle +g_{0 1 }\sqrt{\frac{1}{3}} r^{0 }+g_{2 1 }\sqrt{\frac{1}{6}} r^{2 } }~,\\
F^{1}_{0,0}&=& {\displaystyle +g_{0 1 }\sqrt{\frac{1}{3}} r^{0 }+g_{2 1 }\sqrt{\frac{2}{3}} r^{2 } }~.
\end{array}\right.
\label{psisvampf}
\eeq
In above expressions, 0 and 2 indicate the contributions due to $S$-wave and $D$-wave respectively. Finally, by virtue of Eq.~\eref{decayrate}
\beq
\left.\begin{array}{rcl}
{\displaystyle \frac{d\Gamma}{d\Omega}}
   &\propto & {\displaystyle \sum_{\lambda} |F^1_{0,\lambda}|^2 \cdot |d^{1}_{1 \lambda}(\theta)|^2 }~,\\
   &\propto & {\displaystyle |F^1_{0,1}|^2 (1+\cos^2 \theta) + |F^1_{0,0}|^2  \sin^2 \theta }~.
\end{array}\right.
\label{dkrtpsisv}
\eeq

\subsection{$1^{-}\rightarrow 0^{-} +  0^{-}$}
The typical decays for this mode is $\psi \to \pp$, $\kk$, $\kskl$. From $\eta_J=\eta_s \eta_\sigma (-)^l$, $l$ must equal to 1, that is only $P$-wave is allowed. From $s+\sigma + l -J =$even, then $n=0$. By virtue of Eq.~\eref{xpnofampa}, $A^1_{10}=r$; By virtue of Eq.~\eref{defofampf}, $F^1_{0,0}=g_{10} A^1_{10}=g_{10} r$; finally,
\beq
\left.\begin{array}{rcl}
{\displaystyle \frac{d\Gamma}{d\Omega}}
   &\propto & {\displaystyle \sum_{\lambda} |F^1_{0,0}|^2 \cdot |d^{1}_{\lambda 0}(\theta)|^2 }~,\\
   &\propto & {\displaystyle |F^1_{0,0}|^2  \sin^2 \theta }~.
\end{array}\right.
\label{dkrtpsipp}
\eeq

\subsection{$1^{-}\rightarrow 0^{-} +  1^{-}$}
First, the restriction due to the orbital-angular momentum, $\eta_J=\eta_s \eta_\sigma (-)^l$, leads to $l$=odd. Second, the $lS$-coupling indicates that  only $P$-wave is allowed, i.e., $l=1$. Third, Eq.~\eref{rlnforffn} yields
$$ F^1_{0,1}=-F^1_{0,-1}~, ~~~F^1_{0,0}=-F^1_{0,0}~,$$
so that there is solely one independent invariant helicity-amplitudes, $F^1_{0,1}$. Fourth, since $s+\sigma + l -J =$odd, then $n=1$. By virtue of Eq.~\eref{xpnofampa}, $A^1_{01}(\lambda)=\langle 1 0 1 \lambda |1 \lambda \rangle  w r f^{1}_{\lambda} (\gamma_s)$; By virtue of Eq.~\eref{defofampf},
$F^{1}_{0,\pm 1}= \pm g_{11} \sqrt{1/2} w r$; finally,
\beq
\left.\begin{array}{rcl}
{\displaystyle \frac{d\Gamma}{d\Omega}}
   &\propto & {\displaystyle \sum_{\lambda} |F^1_{0,\lambda}|^2 \cdot |d^{1}_{1 \lambda}(\theta)|^2 }~,\\
   &\propto & {\displaystyle |F^1_{0,1}|^2 (1+\cos^2 \theta) }~.
\end{array}\right.
\label{dkrtpsipv}
\eeq

\subsection{$1^{-}\rightarrow 1^{-} +  1^{+}$}
One of decays for this mode is $\psi \to \phi f_1(1420)$.

The relation $\eta_J=\eta_s \eta_\sigma (-)^l$ leads to $l$=even. Unlike the aforementioned examples, $S$ can take several values, i.e., $S=1 \oplus 1=0,~1,~2$. According to the $lS$-coupling scheme, some possible couplings are listed below
\[
\left.\begin{array}{lll}
  S=0             &  S=1                  &  S=2   \\
 0 \oplus l=l;    & 1 \oplus 0=1;         &  2 \oplus 0=2;  \\
(l=\mbox{even})   & 1 \oplus 2=1,~2,~3;   &  2 \oplus 2=0,~1,~2,~3,~4;   \\
                  & 1 \oplus 4=3,~4,~5;   &  2 \oplus 4=2,~3,~4,~5,~6.
\end{array}\right.
\]
In the light of these results, the possible combinations are $(l,S)=(0,1)$, $(2,1)$, and $(2,2)$. The relation $s+\sigma + l -J =$even, requires $n=0$. Moreover, Eq.~\eref{rlnforffn} indicates that
\[
\left.\begin{array}{ll}
  F^1_{1,1}=-F^1_{-1,-1}~,  &  F^1_{1,0}=-F^1_{-1,0}~,   \\
  F^1_{0,1}=-F^1_{0,-1}~,   &  F^1_{0,0}=-F^1_{0,0}~,
\end{array}\right.
\]
so that there are three independent invariant helicity-amplitudes, viz., $F^1_{1,1}$, $F^1_{1,0}$, and $F^1_{0,1}$. By virtue of Eq.~\eref{xpnofampa},
\beq
\left.\begin{array}{rl}
A^1_{01}(\lambda,\mu) &= {\displaystyle \sqrt{\frac{1}{3}} \langle 1 \lambda 1 -\mu |1 \delta \rangle
w r^0 f^{1}_{\lambda} (\gamma_s) f^{1}_{\mu} (\gamma_\sigma) }~,  \\
A^1_{21}(\lambda,\mu) &= {\displaystyle \sqrt{\frac{5}{3}} \langle 2 0 1 \delta |1 \delta \rangle  \langle 1 \lambda 1 -\mu |1 \delta \rangle
w r^2 f^{1}_{\lambda} (\gamma_s) f^{1}_{\mu} (\gamma_\sigma) }~,  \\
A^1_{22}(\lambda,\mu) &= {\displaystyle \sqrt{\frac{5}{3}} \langle 2 0 2 \delta |1 \delta \rangle  \langle 1 \lambda 1 -\mu |2 \delta \rangle
w r^2 f^{1}_{\lambda} (\gamma_s) f^{1}_{\mu} (\gamma_\sigma) }~.
\end{array}\right.~~
\label{psivpampa}
\eeq
By virtue of Eq.~\eref{defofampf}, the expressions for $F^1_{\lambda,\mu}$'s can be obtained, the more details can refer to the appendix.

\subsection{$1^{-}\rightarrow 1^{-} +  2^{+}$}
One of decays for this mode is $\psi \to \phi f_2(1270)$.

The relation $\eta_J=\eta_s \eta_\sigma (-)^l$ leads to $l$=even, and $S=1 \oplus 2=1,~2,~3$. According to the $lS$-coupling scheme, some possible couplings are listed below
\[
\left.\begin{array}{lll}
  S=1                &  S=2                  &  S=3   \\
 1\oplus 0=1;        & 2\oplus 0=2~;         &  3\oplus 0=3~;  \\
 1\oplus 2=1,~2,~3;  & 2\oplus 2=0,\cdots,4; &  3\oplus 2=1,\cdots,5;   \\
 1\oplus 4=3,~4,~5;  & 2\oplus 4=2,\cdots,6; &  3\oplus 4=1,\cdots,7; \\
                     &                       &  3\oplus 6=3,\cdots,9.
\end{array}\right.
\]
In the light of these results, the possible combinations are $(l,S)=(0,1)$, $(2,1)$, $(2,2)$, $(2,3)$, and $(4,3)$. The relation $s+\sigma + l -J =$even, requires $n=0$. In order to determine the independent $F^1_{\lambda,\mu}$'s, all possibility is listed as follows
\[
F^1_{\lambda,\mu} =
\left(\begin{array}{lllll}
F^1_{2,1}  & F^1_{1,1}  &  F^1_{0,1}  &  F^1_{-1,1}  & F^1_{-2,1}  \\
F^1_{2,0}  & F^1_{1,0}  &  F^1_{0,0}  &  F^1_{-1,0}  & F^1_{-2,0}  \\
F^1_{2,-1} & F^1_{1,-1} &  F^1_{0,-1} &  F^1_{-1,-1} & F^1_{-2,-1}
\end{array}\right)~.
\]
The condition $|\lambda - \nu| \leqslant 1$ vetoes these terms, viz., $F^1_{-1,1}$, $F^1_{-2,0}$, $F^1_{2,0}$, $F^1_{-2,0}$, $F^1_{2,-1}$, and $F^1_{1,-1}$. For the left nine terms, according to Eq.~\eref{rlnforffn},
\[
\left.\begin{array}{ll}
  F^1_{1,1}=F^1_{-1,-1}~,  &  F^1_{1,0}=F^1_{-1,0}~,   \\
  F^1_{0,1}=F^1_{0,-1}~,   &  F^1_{-1,1}=F^1_{1,-1}~,
\end{array}\right.
\]
together  with $F^1_{0,0}$, there are totally five independent invariant helicity-amplitudes.
Then by virtue of Eqs.~\eref{defofampf}, \eref{xpnofampa}, and \eref{decayrate}, the corresponding results can be acquired and the more details can refer to Table~\ref{decayampmmtx} and the appendix.

\begin{table*}[bth]
\caption{\label{decayampmmtx}Various distributions for charmonium decays, that is $\psp$ or $\jpsi$ into final states $M_1$ and $M_2$, where $M_{1,2}=S$, $P$, $V$, $A$, and $T$. The shorthand symbols are defined for the terseness of the table, which include $f_{\lambda \mu} \equiv |F^1_{\lambda, \mu}|^2$,  $\Ctt \equiv 1+\cos^2 \theta$, $\Stt \equiv \sin^2 \theta$, $K_{\alpha} \equiv (\frac{2}{3} \gamma_{\alpha}+\frac{1}{3})~(\alpha=s,\sigma)$,  $K_{\sigma s} \equiv \gamma_{\sigma} \gamma_{s}$, $\kappa = \sqrt{1-2({w_s^2+w_{\sigma}^2})/{w^2}}$, and $g_i$ stands for certain $g_{l S}$.}
\begin{ruledtabular}
\begin{tabular}{cll}
Final state  & $|{\cal M}^{\Gamma}|^2$ for $\psi \to M_1 M_2$ & $|{\cal M}^{\Gamma}|^2$ for $\EE \to \psi \to M_1 M_2$  \\ \hline
SS   & $f_{00} \Stt $                        & $g_1^2 \kappa^2 \Stt $  \\
SP   & $\cdots \cdots $                      & $\cdots \cdots$  \\
SV   & $f_{10} \Ctt + f_{00} \Stt $          & $g_1^2  ( \Ctt + \Stt  )$  \\
SA   & $f_{10} \Ctt $                        & $g_1^2 \kappa^2 \Ctt $  \\
ST   & $f_{10} \Ctt + f_{00} \Stt $
& $ g_1^2 \kappa^2 ( \frac{3}{10} \gamma_{\sigma}^2 \Ctt +  \frac{2}{5} K_{\sigma}^2 \Stt )$  \\
PP   & $f_{00} \Stt $                        & $ g_1^2 \kappa^2 \Stt $   \\
PV   & $f_{10} \Ctt $                        & $\frac{1}{2} g_1^2 \kappa^2 \Ctt $  \\
PA   & $f_{10} \Ctt + f_{00} \Stt $          & $\frac{1}{3} g_1^2 ( \Ctt + \Stt )$ \\
PT   & $f_{10} \Ctt  $                       & $\frac{1}{2} g_1^2 \kappa^4 \Ctt $   \\
VV   &  $ \left.\begin{array}{l} [f_{11}+\frac{1}{2}f_{00}] \Stt \\
+\frac{1}{2} [f_{10}+f_{01}] \Ctt \end{array}\right. $
& $ \left.\begin{array}{l}
 \kappa^2 \Stt \cdot [\frac{1}{3} (1+\frac{1}{2}K_{\sigma s}^2 ) \cdot g_1^2
 + \frac{1}{5} (1+2 K_{\sigma s}^2 )\cdot g_3^2
 - \frac{2}{3\sqrt{5}} (1-\frac{1}{2} K_{\sigma s}^2 ) \cdot g_1 g_3 ]  \\
 +\frac{1}{2} \kappa^2 \Ctt \cdot [ (\gamma_{\sigma}^2+\gamma_{s}^2)\cdot (\frac{1}{4} g_2^2 + \frac{3}{20} g_3^2)
 +\sqrt{\frac{3}{20}} (\gamma_{\sigma}^2-\gamma_{s}^2) \cdot g_2 g_3]  \end{array}\right. $\\
VA   &  $ \left.\begin{array}{l} f_{11} \Stt \\
+\frac{1}{2} [f_{10}+f_{01}] \Ctt \end{array}\right. $
& $ \frac{1}{6} \cdot [ \Stt + \Ctt \cdot (\gamma_{\sigma}^2+\gamma_{s}^2) ] \cdot g_1^2  $  \\
VT   &  $ \left.\begin{array}{l} [f_{11}+\frac{1}{2}f_{00}] \Stt \\
+\frac{1}{2} [f_{12}+f_{10}+f_{01}] \Ctt \end{array}\right. $
& $ \left.\begin{array}{l}
 \Stt \cdot ( \frac{1}{10} \gamma_{\sigma}^2 + \frac{2}{15} \gamma_s^2 K_{\sigma}^2 ) \cdot g_1^2  \\
+\frac{1}{2} \Ctt \cdot ( \frac{1}{5} + \frac{1}{30}K_{\sigma}^2 + \frac{1}{10 }K_{s}^2 ) \cdot g_1^2
\end{array}\right. $ \\
AA   &  $ \left.\begin{array}{l} [f_{11}+\frac{1}{2}f_{00}] \Stt \\
+\frac{1}{2} [f_{10}+f_{01}] \Ctt \end{array}\right. $
& $ \left.\begin{array}{l}
 \kappa^2 \Stt \cdot [\frac{1}{3} (1+\frac{1}{2}\gamma_s^2 ) \cdot g_1^2
 + \frac{1}{15} (1+2 \gamma_s^2 ) \cdot g_3^2+\frac{2}{3\sqrt{5}} (1+\frac{1}{2}\gamma_s^2 )\cdot g_1 g_3] \\
 +\frac{1}{2} \kappa^2 \Ctt \cdot [ (\gamma_{\sigma}^2+\gamma_{s}^2) \cdot (\frac{1}{4}  g_2^2
 + \frac{3}{20} g_3^2 )
 - \frac{1}{30} (\gamma_{\sigma}^2-\gamma_{s}^2)\cdot  g_2 g_3 ] \\
\end{array}\right. $  \\
AT  &  $ \left.\begin{array}{l}  f_{11} \Stt \\
+\frac{1}{2} [f_{12}+f_{10}+f_{01}] \Ctt \end{array}\right. $
& $ \frac{1}{5} \kappa^2 \Stt g_2^2 \gamma_{\sigma}^2 +
\frac{1}{2} \kappa^2 \Ctt \cdot \left\{ \begin{array}{l}
[\frac{1}{5} + \frac{1}{20} K_{\sigma}^2+ +\frac{3}{20} K_{\sigma s}^2]\cdot g_1^2  \\
+[\frac{1}{10} + \frac{3}{20} K_{\sigma}^2+ +\frac{1}{20} K_{\sigma s}^2]\cdot g_2^2 \\
+[-\frac{\sqrt{2}}{5} + \frac{\sqrt{3}}{10} K_{\sigma}^2+ +\frac{\sqrt{3}}{10} K_{\sigma s}^2]\cdot g_1 g_2
\end{array} \right\} $  \\
TT  &  $ \left.\begin{array}{l}
[f_{22}+f_{11}+\frac{1}{2}f_{00}] \Stt \\
+\frac{1}{2} [f_{12}+f_{21}+f_{10}+f_{01}] \Ctt
\end{array}\right. $
& $ \left.\begin{array}{l}
\kappa^2 \Stt \cdot \left\{ \begin{array}{l}
\frac{1}{5} (1+K^2_{\sigma s}+\frac{1}{2} K_s^2 K_{\sigma}^2) \cdot g_1^2  \\
+\frac{1}{35} (4+K^2_{\sigma s}+ 2 K_s^2 K_{\sigma}^2) \cdot g_3^2 \\
- \frac{2}{5 \sqrt{7}}  (2+K^2_{\sigma s} - K_s^2 K_{\sigma}^2) \cdot g_1 g_3
\end{array} \right\}  \\
+\frac{1}{2} \kappa^2 \Ctt \cdot \left\{ \begin{array}{l}
[\frac{1}{10} (\gamma_{\sigma}^2+\gamma_{s}^2)+\frac{3}{20} (\gamma_{\sigma}^2 K_s^2+\gamma_{s}^2 K_{\sigma}^2)]\cdot g_2^2  \\
+[\frac{9}{70} (\gamma_{\sigma}^2+\gamma_{s}^2)+\frac{3}{140} (\gamma_{\sigma}^2 K_s^2+\gamma_{s}^2 K_{\sigma}^2)]\cdot g_3^2 \\
+ 2 [\frac{3}{10 \sqrt{7}}(\gamma_{\sigma}^2-\gamma_{s}^2)+\frac{3}{20\sqrt{7}} (\gamma_{s}^2 K_{\sigma}^2-\gamma_{\sigma}^2 K_s^2)]\cdot g_2 g_3
\end{array} \right\}
\end{array}\right.
$  \\
\end{tabular}
\end{ruledtabular}
\end{table*}

\section{Cross sections}\label{xct_krxns}
Based on field theory~\cite{weinberg1995,dewitand1986}, the cross section for the process $a+b \to V \to f$ reads
\beq
\left.\begin{array}{l}
\sigma (a+b\to V \to f) ={\displaystyle \frac{(2S_V+1)}{(2S_a+1)(2S_b+1)}  } \\
        \cdot {\displaystyle \frac{16\pi m_V^4}{\lambda (m^2_V,m^2_a,m^2_b)} }
         \cdot {\displaystyle  \frac{\Gamma (V\to f)\cdot \Gamma (\bar{V} \to \bar{a}+\bar{b})}{(s-m^2_V)^2+m^2_V \Gamma^2 (V\to all)} }~,
\end{array}\right.
\label{totxnforabvf}
\eeq
where $S_V$, $S_a$, and $S_b$ denote the spin of particles $V$, $a$, and $b$, respectively; $m_V$, $m_a$, and $m_b$ are their corresponding masses; $\Gamma (V\to f)$, $\Gamma (\bar{V} \to \bar{a}+\bar{b})$, and $\Gamma (V\to all)$ (also denoted as $\Gamma_t$) are the partial and total decay rates in the rest frame of the particle $V$; $\lambda (x,y,z)$ is a function with the definition $\lambda (x,y,z) \equiv (x-y-z)^2-4 yz$.

Here the inverse process of $a+b \to V$ is the process $\bar{V} \to \bar{a}+\bar{b}$, where $\bar{V}$, $\bar{a}$, and $\bar{b}$ are the antiparticles associated with $V$, $a$, and $b$, which may again be the same particles. This is just the case for the electron and positron collision experiment.

The annihilation processes of $e^+$-$e^-$ take place through the conversion of the pair into a virtual photon of mass equal to the total center-of-mass (CM) energy. The photon converts into the final products. These reactions proceed through a state of well-defined quantum numbers, and as consequence the possible initial and final states are essentially limited. The interaction of the final particles with the virtual photon is directly measured in the experiment. Unstable particles with the same quantum numbers as the intermediate photon can be produced singly as resonant states that soon after decay. At the appropriate energy there would appear resonance peaks in the production cross section for the final decay products, which is just the case described by Eq.~\eref{totxnforabvf}. Specially speaking, if $V=\psi$, $\Gamma (\psi \to \EE)$ is denoted by $\Gamma_e$ and $\Gamma (\psi \to f)$ by $\Gamma_f$, moreover, if the masses of electron and positron are neglected, the cross section $\sigma ( e^+ + e^- \to \psi \to f)$ (denoted by $\sigma_f$) can be recast as

\beq
\sigma_f (s) = \frac{12 \pi \cdot \Gamma_e \Gamma_f}{(s-m^2_V)^2+m^2_V \Gamma^2_t} ~.
\label{sigmafforeepsif}
\eeq

In order to utilize the results obtained in the previous section, we are to derive the above relation through another way. Start with the differential cross section~\cite{quangpham}
\beq
\frac{d\sigma (a+b \to V \to X+Y)}{d\Omega} = \frac{|{\cal M}^{tot}|^2}{64 \pi^2 s } \cdot \frac{|q_f|}{|q_i|}~,
\label{dffxnforabvxy}
\eeq
where $\sqrt{s}$ is the CM energy of particle $V$ and $q_f$ ($q_i$) is the momentum of either the $X$ ($a$) or the $Y$ ($b$). The total decay amplitude can be constructed as follows
\beq
{\cal M}^{tot} = {\cal M}^{D} (V\to X+Y) \cdot D(s) \cdot {\cal M}^{P} (a+b\to V) ~,
\label{mtandmdmp}
\eeq
with
\beq
D(s) = \frac{\xi}{s-m^2_V+ i m_V \Gamma_t} ~,
\label{pgtofres}
\eeq
which is the propagator of resonance $V$. The parameter $\xi$ will be fixed later.

By virtue of the principle of detailed balancing~\cite{landau1977}, ${\cal M}^{P} (a+b\to V)={\cal M}^{\bar{D}} (\bar{V} \to \bar{a}+\bar{b})$. Now turn to the electron and positron collision experiment, since $e^+$ and $e^-$ are defined along the $z$ axis, there is no directionally dependent variable for ${\cal M}^{\bar{D}} (\psi \to \EE)$, so after the integral of directional variable, it obtains
\beq
\Gamma_e \equiv \Gamma (\psi \to \EE) = \frac{|{\cal M}^{\bar{D}}|^2}{8 \pi} \cdot \frac{|q_i|}{m^2_\psi}~.
\label{dkgmaee}
\eeq
If we define $\Gamma_f$ as
\beq
\Gamma_f \equiv \Gamma (\psi \to f) =\int d\Omega \frac{|{\cal M}^{D}|^2}{32 \pi^2} \cdot \frac{|q_f|}{m^2_\psi}~,
\label{dkgmaf}
\eeq
and put all these together, then utilizing Eq.~\eref{dffxnforabvxy}, at the same time neglecting the masses of electron and positron, it obtains eventually
\beq
\sigma_f (s) = \frac{4\pi \xi^2 \cdot \Gamma_e \Gamma_f}{(s-m^2_\psi)^2+m^2_\psi \Gamma^2_t} ~.
\label{sigmaseond}
\eeq
Comparing Eq.~\eref{sigmafforeepsif} with Eq.~\eref{sigmaseond}, notice that $V=\psi$ herein, $\xi$ is fixed to be $\sqrt{3}$. Here the crucial point lies in the fact that ${\cal M}^{D}$ in Eq.~\eref{dkgmaf} is just the ${\cal M}^{\Gamma}$ in Eq.~\eref{decayrate}, therefore, the results for decay rates can be adopted for the calculations of cross sections.

The collision cross sections are calculated and results are tabulated in Table~\ref{decayampmmtx}. The information involving $F^J_{\lambda \nu}$ has been collected in the appendix, and only the lowest order of $r^l$ is adopted in the calculation.

A remark is in order here. Table~\ref{decayampmmtx} displays the distributions of charmonium decays, where the charmonium, such as $\psp$ or $\jpsi$, is produced in an $\EE$ collider. The spin projections in this case can only be either $+1$ or $-1$ along the direction of the $e^+$ or $e^-$ beams~\cite{halzen1984}, so that the zero component for $d$-function is forbidden. This requirement decreases the number of independent $F^J_{\lambda \nu}$'s.
However, if the charmonium is from the decay product, say, $\jpsi$ comes from $\EE \to \psp \to \pi^+\pi^- \jpsi$, then in the rest frame of $\jpsi$ particle, the angular distribution of the two-body decay is a trivial event distribution. But the distribution may implicate more than one form factor. Let us consider the SV mode concretely.
For the cascade decay $\psp \to \pi^+\pi^- \jpsi$, then $\jpsi \to \phi f_0(980)$, in the rest frame of $\jpsi$ particle, the angular distribution reads
\beq
\frac{d\sigma}{d\Omega} \propto |F^1_{0,1}|^2 + |F^1_{0,0}|^2 ~,
\label{dsbunjpssv}
\eeq
which can be obtained from the formula~\eref{mtxsqmgamma}, where $M$ can be $\pm$ and $0$; instead of the formula~\eref{dkrtpsisv}, where $M$ can solely be $\pm$. Obviously, there are two form factors, that is $F^1_{0,1}$ and $F^1_{0,0}$, for this event distribution.

If we combine this distribution with that from an $\EE$ collider, we can acquire more useful information.
According to Eq.~\eref{dkrtpsisv} the decay rate reads
\beq
\frac{d\Gamma}{d\Omega} \propto |F^1_{0,1}|^2 (1+\cos^2 \theta) + |F^1_{0,0}|^2  \sin^2 \theta~,
\label{dkrtpsisv2}
\eeq
If we introduce the rate $\alpha_f$ as follows
\beq
\alpha_f =\frac{|F^1_{0,1}|^2-|F^1_{0,0}|^2}{|F^1_{0,1}|^2+|F^1_{0,0}|^2}~,
\label{ratealf}
\eeq
then Eq.~\eref{dkrtpsisv2} can be recast as
\beq
\frac{d\Gamma}{d\Omega} \propto 1 + \alpha_f \cos^2 \theta~,
\label{dkrtpsisv2a}
\eeq
by virtue of which $\alpha_f$ can be determined, and in turn it can be obtained
\beq
\frac{|F^1_{0,0}|^2}{|F^1_{0,1}|^2}=\frac{1-\alpha_f}{1+\alpha_f}~.
\label{rateff}
\eeq
Such an analysis can help us obtain the relative relation among form factors.

\section{Summary}\label{xct_sum}
In this paper, the invariant helicity amplitudes for various charmonium decay are studied. The general formulas are summarized and a number of illustrative examples is worked out to expound and incarnate the theoretical ideas and general criteria.

The relation between the decay rate and the collision cross section is derived, so the results of the decay rate can be utilized for the cross section calculation. The electromagnetic cross sections of charmonium decay in an $\EE$ collider are valuated for fifteen decay modes. Combining these results with the systematical parametrization scheme, the colossal charmonium data samples taken in BESIII detector could be analyzed systematically in the future, which will shed the light on the dynamics of charmonium decay .

\section*{Acknowledgment}
The author would like to thank Prof. C.Z.~yuan and Prof. G.~Li for helpful discussions and thanks is also due to Prof. G.~Li for program calculation.

This work is supported in part by National Key Research and Development Program of China under Contracts No.~2023YFA1606000 and No.~2023YFA1606003.

\appendix
\begin{appendix}
\section{}
This appendix summarizes, in toto, the calculation results of $F^J_{\lambda \nu}$ for the fifteen modes of charmonium decay. The factor $w$ has been normalized out.

\begin{enumerate}
\item {$1^{-}\rightarrow 0^{+} +  0^{+}$}
\beq
\left.\begin{array}{rcl}
F^{1}_{0,0}&=& +g_{1 0 }\sqrt{\frac{1}{1}} r^{1 }
\end{array}\right.
\eeq

\item {$1^{-}\rightarrow 0^{+} +  0^{-}$} \\
\[
\mbox{This mode is forbidden. }
\]

\item {$1^{-}\rightarrow 0^{+} +  1^{-}$} \\
\beq
\left.\begin{array}{rcl}
F^{1}_{0,1}&=& +g_{0 1 }\sqrt{\frac{1}{3}} r^{0 }+g_{2 1 }\sqrt{\frac{1}{6}} r^{2 } \\
F^{1}_{0,0}&=& +g_{0 1 }\sqrt{\frac{1}{3}} r^{0 }-g_{2 1 }\sqrt{\frac{2}{3}} r^{2 }
\end{array}\right.
\eeq

\item {$1^{-}\rightarrow 0^{+} +  1^{+}$} \\
\beq
\left.\begin{array}{rcl}
F^{1}_{0,1}&=& +g_{1 1 }\sqrt{\frac{1}{2}} r^{1 }
\end{array}\right.
\eeq

\item {$1^{-}\rightarrow 0^{+} +  2^{+}$} \\
\beq
\left.\begin{array}{rcl}
F^{1}_{0,1}&=& -g_{1 2 }\sqrt{\frac{3}{10}} r^{1 } \gamma_\sigma -g_{3 2 }\sqrt{\frac{1}{5}} r^{3 } \gamma_\sigma \\
F^{1}_{0,0}&=& -g_{1 2 }\sqrt{\frac{2}{5}} r^{1 }  (\frac{2}{3}\gamma^2_\sigma+\frac{1}{3})
 +g_{3 2 }\sqrt{\frac{3}{5}} r^{3 } (\frac{2}{3}\gamma^2_\sigma+\frac{1}{3})
\end{array}\right.
\eeq

\item {$1^{-}\rightarrow 0^{-} +  0^{-}$} \\
\beq
\left.\begin{array}{rcl}
F^{1}_{0,0}&=& +g_{1 0}\sqrt{\frac{1}{1}} r^{1 }
\end{array}\right.
\eeq

\item {$1^{-}\rightarrow 0^{-} +  1^{-}$} \\
\beq
\left.\begin{array}{rcl}
F^{1}_{0,1}&=& +g_{1 1 }\sqrt{\frac{1}{2}} r^{1 }
\end{array}\right.
\eeq

\item {$1^{-}\rightarrow 0^{-} +  1^{+}$} \\
\beq
\left.\begin{array}{rcl}
F^{1}_{0,1}&=& +g_{0 1 }\sqrt{\frac{1}{3}} r^{0 }+g_{2 1 }\sqrt{\frac{1}{6}} r^{2 } \\
F^{1}_{0,0}&=& +g_{0 1 }\sqrt{\frac{1}{3}} r^{0 }-g_{2 1 }\sqrt{\frac{2}{3}} r^{2 }
\end{array}\right.
\eeq

\item {$1^{-}\rightarrow 0^{-} +  2^{+}$} \\
\beq
\left.\begin{array}{rcl}
F^{1}_{0,1}&=& -g_{2 2 }\sqrt{\frac{1}{2}} r^{2 }
\end{array}\right.
\eeq

\item {$1^{-}\rightarrow 1^{-} +  1^{-}$} \\
\beq
\left.\begin{array}{rcl}
F^{1}_{1,1}&=& +g_{1 0}\sqrt{\frac{1}{3}} r^{1 } \\
           & & -g_{1 2}\sqrt{\frac{1}{15}} r^{1 }+g_{3 2}\sqrt{\frac{1}{10}} r^{3 } \\
F^{1}_{1,0}&=& -g_{1 1}\sqrt{\frac{1}{4}} r^{1 }\gamma_\sigma  \\
           & & -g_{1 2}\sqrt{\frac{3}{20}} r^{1 } \gamma_\sigma-g_{3 2}\sqrt{\frac{1}{10}} r^{3 } \gamma_\sigma \\
F^{1}_{0,1}&=& +g_{1 1}\sqrt{\frac{1}{4}} r^{1 }\gamma_s \\
           & & -g_{1 2}\sqrt{\frac{3}{20}} r^{1 } \gamma_s-g_{3 2}\sqrt{\frac{1}{10}} r^{3 } \gamma_s \\
F^{1}_{0,0}&=& -g_{1 0}\sqrt{\frac{1}{3}} r^{1 }\gamma_s  \gamma_\sigma\\
           & & -g_{1 2}\sqrt{\frac{4}{15}} r^{1 } \gamma_s \gamma_\sigma
               +g_{3 2}\sqrt{\frac{2}{5}} r^{3 } \gamma_s \gamma_\sigma
\end{array}\right.
\eeq

\item {$1^{-}\rightarrow 1^{-} +  1^{+}$} \\
\beq
\left.\begin{array}{rcl}
F^{1}_{1,1}&=& +g_{0 1}\sqrt{\frac{1}{6}} r^{0 }-g_{2 1}\sqrt{\frac{1}{3}} r^{2 } \\
F^{1}_{1,0}&=& +g_{0 1}\sqrt{\frac{1}{6}} r^{0 }\gamma_\sigma  \\
           & & +g_{2 1}\sqrt{\frac{1}{12}} r^{2 }\gamma_\sigma+g_{2 2}\sqrt{\frac{1}{4}} r^{2 } \gamma_\sigma \\
F^{1}_{0,1}&=& +g_{0 1}\sqrt{\frac{1}{6}} r^{0 }\gamma_s \\
           & & +g_{2 1}\sqrt{\frac{1}{12}} r^{2 } \gamma_s-g_{2 2}\sqrt{\frac{1}{4}} r^{2 } \gamma_s
\end{array}\right.
\eeq

\item {$1^{-}\rightarrow 1^{-} +  2^{+}$} \\
\beq
\left.\begin{array}{rcl}
F^{1}_{1,2}&=& +g_{0 1}\sqrt{\frac{1}{5}} r^{0 } \\
           & & +g_{2 1}\sqrt{\frac{1}{10}} r^{2 }-g_{2 2}\sqrt{\frac{1}{6}} r^{2 } \\
           & & +g_{2 3}\sqrt{\frac{2}{105}} r^{2 }+g_{4 3 }\sqrt{\frac{1}{70}} r^{4 } \\
F^{1}_{1,1}&=& +g_{0 1}\sqrt{\frac{1}{10}} r^{0 } \gamma_\sigma-g_{2 1}\sqrt{\frac{1}{5}} r^{2 }\gamma_\sigma\\
           & & +g_{2 3}\sqrt{\frac{3}{35}} r^{2 }\gamma_\sigma-g_{4 3}\sqrt{\frac{4}{35}} r^{4 }\gamma_\sigma \\
F^{1}_{1,0}&=& +g_{0 1}\sqrt{\frac{1}{30}} r^{0 }(\frac{2}{3}\gamma^2_\sigma+\frac{1}{3}) \\
           & & +g_{2 1}\sqrt{\frac{1}{60}} r^{2 } (\frac{2}{3}\gamma^2_\sigma+\frac{1}{3})
               +g_{2 2}\sqrt{\frac{1}{4}} r^{2 } (\frac{2}{3}\gamma^2_\sigma+\frac{1}{3}) \\
           & & +g_{2 3}\sqrt{\frac{4}{35}} r^{2 } (\frac{2}{3}\gamma^2_\sigma+\frac{1}{3})
               +g_{4 3}\sqrt{\frac{3}{35}} r^{4 } (\frac{2}{3}\gamma^2_\sigma+\frac{1}{3})  \\
F^{1}_{0,1}&=& -g_{0 1}\sqrt{\frac{1}{10}} r^{0 } \gamma_s \gamma_\sigma \\
           & & -g_{2 1}\sqrt{\frac{1}{20}} r^{2 } \gamma_s \gamma_\sigma-g_{2 2}\sqrt{\frac{1}{12}} r^{2 } \gamma_s \gamma_\sigma\\
           & & +g_{2 3}\sqrt{\frac{16}{105}} r^{2 } \gamma_s \gamma_\sigma+g_{4 3}\sqrt{\frac{4}{35}} r^{4 } \gamma_s \gamma_\sigma\\
F^{1}_{0,0}&=& -g_{0 1}\sqrt{\frac{2}{15}} r^{0 }\gamma_s (\frac{2}{3}\gamma^2_\sigma+\frac{1}{3})
               +g_{2 1}\sqrt{\frac{4}{15}} r^{2 }\gamma_s (\frac{2}{3}\gamma^2_\sigma+\frac{1}{3}) \\
           & & +g_{2 3}\sqrt{\frac{9}{35}} r^{2 }\gamma_s (\frac{2}{3}\gamma^2_\sigma+\frac{1}{3})
               -g_{4 3}\sqrt{\frac{12}{35}} r^{4 }\gamma_s (\frac{2}{3}\gamma^2_\sigma+\frac{1}{3})
\end{array}\right.
\eeq

\item {$1^{-}\rightarrow 1^{+} +  1^{+}$} \\
\beq
\left.\begin{array}{rcl}
F^{1}_{1,1}&=& +g_{1 0}\sqrt{\frac{1}{3}} r^{1 } \\
           & & -g_{1 2}\sqrt{\frac{1}{15}} r^{1 }+g_{3 2}\sqrt{\frac{1}{10}} r^{3 } \\
F^{1}_{1,0}&=& -g_{1 1}\sqrt{\frac{1}{4}} r^{1 }\gamma_\sigma  \\
           & & -g_{1 2}\sqrt{\frac{3}{20}} r^{1 } \gamma_\sigma-g_{3 2}\sqrt{\frac{1}{10}} r^{3 } \gamma_\sigma \\
F^{1}_{0,1}&=& +g_{1 1}\sqrt{\frac{1}{4}} r^{1 }\gamma_s \\
           & & -g_{1 2}\sqrt{\frac{3}{20}} r^{1 } \gamma_s-g_{3 2}\sqrt{\frac{1}{10}} r^{3 } \gamma_s \\
F^{1}_{0,0}&=& -g_{1 0}\sqrt{\frac{1}{3}} r^{1 }\gamma_s \gamma_\sigma \\
           & & -g_{1 2}\sqrt{\frac{4}{15}} r^{1 } \gamma_s \gamma_\sigma
               +g_{3 2}\sqrt{\frac{2}{5}} r^{3 } \gamma_s \gamma_\sigma
\end{array}\right.
\eeq

\item {$1^{-}\rightarrow 1^{+} +  2^{+}$} \\
\beq
\left.\begin{array}{rcl}
F^{1}_{1,2}&=& +g_{1 1}\sqrt{\frac{3}{10}} r^{1 } -g_{1 2}\sqrt{\frac{1}{10}} r^{1 }\\
           & & -g_{3 2}\sqrt{\frac{1}{15}} r^{3 } +g_{3 3 }\sqrt{\frac{1}{30}} r^{3 } \\
F^{1}_{1,1}&=& -g_{1 2}\sqrt{\frac{1}{5}} r^{1 } \gamma_\sigma+g_{3 2}\sqrt{\frac{3}{10}} r^{3 }\gamma_\sigma\\
F^{1}_{1,0}&=& -g_{1 1}\sqrt{\frac{1}{20}} r^{1 }(\frac{2}{3}\gamma^2_\sigma+\frac{1}{3})
               -g_{1 2}\sqrt{\frac{3}{20}} r^{1 } (\frac{2}{3}\gamma^2_\sigma+\frac{1}{3}) \\
           & & -g_{3 2}\sqrt{\frac{1}{10}} r^{3 } (\frac{2}{3}\gamma^2_\sigma+\frac{1}{3})
               -g_{3 3}\sqrt{\frac{1}{5}} r^{3 } (\frac{2}{3}\gamma^2_\sigma+\frac{1}{3})  \\
F^{1}_{0,1}&=& -g_{1 1}\sqrt{\frac{3}{20}} r^{1 } \gamma_s \gamma_\sigma -g_{1 2}\sqrt{\frac{1}{20}} r^{1 } \gamma_s \gamma_\sigma \\
           & & -g_{3 2}\sqrt{\frac{1}{30}} r^{3 } \gamma_s \gamma_\sigma+g_{3 3}\sqrt{\frac{4}{15}} r^{3 } \gamma_s \gamma_\sigma
\end{array}\right.
\eeq

\item {$1^{-}\rightarrow 2^{+} +  2^{+}$} \\
\beq
\left.\begin{array}{rcl}
F^{1}_{2,2}&=& +g_{1 0}\sqrt{\frac{1}{5}} r^{1 } -g_{1 2}\sqrt{\frac{4}{35}} r^{1 }  \\
           & & +g_{3 2}\sqrt{\frac{6}{35}} r^{3 } -g_{3 4}\sqrt{\frac{2}{315}} r^{3 } +g_{5 4}\sqrt{\frac{1}{126}} r^{5 } \\
F^{1}_{2,1}&=& -g_{1 1}\sqrt{\frac{1}{10}} r^{1 } \gamma_\sigma -g_{1 2}\sqrt{\frac{9 }{70 }} r^{1 } \gamma_\sigma \\
           & & -g_{3 2}\sqrt{\frac{3}{35}} r^{3 } \gamma_\sigma -g_{3 3}\sqrt{\frac{3 }{20 }} r^{3 } \gamma_\sigma  \\
           & & -g_{3 4}\sqrt{\frac{5}{252}} r^{3 } \gamma_\sigma -g_{5 4}\sqrt{\frac{1 }{63 }} r^{5 }   \gamma_\sigma \\
F^{1}_{1,2}&=& +g_{1 1}\sqrt{\frac{1}{10}} r^{1 }\gamma_s -g_{1 2}\sqrt{\frac{9 }{70 }} r^{1 }\gamma_s \\
           & & -g_{3 2}\sqrt{\frac{3}{35}} r^{3 }\gamma_s +g_{3 3}\sqrt{\frac{3 }{20 }} r^{3 }\gamma_s \\
           & & -g_{3 4}\sqrt{\frac{5}{252}} r^{3 }\gamma_s -g_{5 4}\sqrt{\frac{1 }{63 }} r^{5 }\gamma_s \\
F^{1}_{1,1}&=& -g_{1 0}\sqrt{\frac{1}{5}} r^{1 }\gamma_s \gamma_\sigma
               -g_{1 2}\sqrt{\frac{1}{35}} r^{1 }\gamma_s \gamma_\sigma  \\
           & & +g_{3 2}\sqrt{\frac{3}{70}} r^{3 }\gamma_s  \gamma_\sigma
               -g_{3 4}\sqrt{\frac{32}{315}} r^{3 }\gamma_s \gamma_\sigma \\
           & & +g_{5 4}\sqrt{\frac{8}{63}} r^{5 }\gamma_s  \gamma_\sigma \\
F^{1}_{1,0}&=& +g_{1 1}\sqrt{\frac{3}{20}} r^{1 }\gamma_s  (\frac{2}{3}\gamma^2_\sigma+\frac{1}{3})
               +g_{1 2}\sqrt{\frac{3}{140}} r^{1 }\gamma_s  (\frac{2}{3}\gamma^2_\sigma+\frac{1}{3}) \\
           & & +g_{3 2}\sqrt{\frac{1}{70}} r^{3 }\gamma_s  (\frac{2}{3}\gamma^2_\sigma+\frac{1}{3})
               -g_{3 3}\sqrt{\frac{1}{10}} r^{3 }\gamma_s  (\frac{2}{3}\gamma^2_\sigma+\frac{1}{3}) \\
           & & -g_{3 4}\sqrt{\frac{5}{42}} r^{3 }\gamma_s  (\frac{2}{3}\gamma^2_\sigma+\frac{1}{3})
               -g_{5 4}\sqrt{\frac{2}{21}} r^{5 }\gamma_s (\frac{2}{3}\gamma^2_\sigma+\frac{1}{3}) \\
F^{1}_{0,1}&=& -g_{1 1}\sqrt{\frac{3}{20}} r^{1 }(\frac{2}{3}\gamma^2_s+\frac{1}{3})  \gamma_\sigma
               +g_{1 2}\sqrt{\frac{3}{140}} r^{1 }(\frac{2}{3}\gamma^2_s+\frac{1}{3}) \gamma_\sigma \\
           & & +g_{3 2}\sqrt{\frac{1}{70}} r^{3 }(\frac{2}{3}\gamma^2_s+\frac{1}{3})  \gamma_\sigma
               +g_{3 3}\sqrt{\frac{1}{10}} r^{3 }(\frac{2}{3}\gamma^2_s+\frac{1}{3})  \gamma_\sigma \\
           & & -g_{3 4}\sqrt{\frac{5}{42}} r^{3 }(\frac{2}{3}\gamma^2_s+\frac{1}{3})  \gamma_\sigma
               -g_{5 4}\sqrt{\frac{2}{21}} r^{5 }(\frac{2}{3}\gamma^2_s+\frac{1}{3})  \gamma_\sigma \\
F^{1}_{0,0}&=& +g_{1 0}\sqrt{\frac{1}{5}} r^{1 }(\frac{2}{3}\gamma^2_s+\frac{1}{3}) (\frac{2}{3}\gamma^2_\sigma+\frac{1}{3}) \\
           & & +g_{1 2}\sqrt{\frac{4}{35}} r^{1 }(\frac{2}{3}\gamma^2_s+\frac{1}{3}) (\frac{2}{3}\gamma^2_\sigma+\frac{1}{3}) \\
           & & -g_{3 2}\sqrt{\frac{6}{35}} r^{3 }(\frac{2}{3}\gamma^2_s+\frac{1}{3}) (\frac{2}{3}\gamma^2_\sigma+\frac{1}{3}) \\
           & & -g_{3 4}\sqrt{\frac{8}{35}} r^{3 }(\frac{2}{3}\gamma^2_s+\frac{1}{3})  (\frac{2}{3}\gamma^2_\sigma+\frac{1}{3}) \\
           & & +g_{5 4}\sqrt{\frac{2}{7}} r^{5 }(\frac{2}{3}\gamma^2_s+\frac{1}{3})  (\frac{2}{3}\gamma^2_\sigma+\frac{1}{3})
\end{array}\right.
\eeq

\end{enumerate}

\end{appendix}

\end{document}